\documentclass[review]{elsarticle}

\usepackage{lineno,hyperref,amsmath, float}

\modulolinenumbers[5]

\journal{Future Generation Computer Systems}

\begin{document}

\begin{frontmatter}

\title{Blockchain Tree as Solution for Distributed Storage of Personal ID Data and Document Access Control}

\author[First,Second]{Sergii Kushch\corref{cor1}}
\ead{kushch@yaros.co}

\author[Third]{Yurii Baryshev}
\ead{yuriy.baryshev@gmail.com}

\author[First]{Silvio Ranise}
\ead{ranise@fbk.eu}

\cortext[cor1]{Corresponding author.}

\address[First]{Security and Trust research unit, Bruno Kessler Foundation,\\
Via Sommarive, 18, Povo, TN, 38123, Italy}
\address[Second]{University of Limerick, Limerick, V94 T9PX, Ireland}
\address[Third]{Information Protection Department, Vinnytsia National Technical University,\\
95 Khmelnytske shose, Vinnytsia, 21021, Ukraine}

\begin{abstract}
This paper introduces a new method of Blockchain formation for reliable storage of personal data of ID-card holders. In particular, the model of the information system is presented, the new structure of smart ID-cards and information on these cards are proposed. The new structure of Blockchain - "Blockchain Tree" allows not only to store information from ID-cards but also to increase the level of security and access control to this information. The proposed Subchains system allows to integrate Blockchain of the lower level to Blockchain of the higher level, allowing to create a multilevel protected system.
\end{abstract}


\end{frontmatter}


\section{Introduction}

The use of information resources and government services in the modern world implies unambiguous user identification. To this end, many states create so-called digital personalities. The most famous example of such a system in the EU is Estonia. In this country, every citizen can get not only the usual ID-card for the EU countries but also the mobile-ID. Using their digital ID, an ID-holder may get an online access to most government services, remotely open a bank account, register a company, make an appointment for a doctor, etc.\\
The physical medium of such an identifier is an ID card \cite{BSI24}, \cite{Mach25}, \cite{Doc26}, \cite{Carta27}, \cite{Trub28}. All countries of the European Union issue national ID-cards to citizens and residents.\\
The chip of the card stores information about its owner: full name, gender, national identification number, fingerprints, cryptographic keys and certificates. A card holder has the right to use the ID-card as an identity card for travel through the territory of the European Union and for crossing its external borders both for entry and exit from the countries of the European Union, the European Economic Area, including Iceland, Norway and Switzerland.\\
We consider the issue of this card’s security, because a single card contains a key to all personal data of a citizen.\\
Firstly, the chip on a smart card is a sufficiently protected microcomputer that has a microprocessor, a cryptographic co-processor, and some memory (flash or EEPROM). Unlike a standard micro-controller, an access to the memory of a smart card is strictly controlled by the processor. Thus, both reading and recording of the data are regulated by the software of the card itself. Moreover, chip manufacturers are taking measures to prevent unauthorized access (copying all the memory, reprogramming) to the card at electronic and physical levels.\\
Secondly, all the data in the card is encrypted (in the Estonian map 2048 bit encryption is used).
Thirdly, the downloaded applications (applets) cannot be read from the card by anyone, including the cardholder (an applet can only be erased and a new one should be written instead of the erased one on its place). You can only pass a command to the applet and get a response.\\
Fourth, you need a PIN-code for the card, and sometimes two: the first one for authorization and the second one for confirmation of operations. \\
However, despite all the advantages of such protection, the system is not secure from unauthorized changes of personal information from the inside (for example, due to server hacking, etc.), from creation of fake digital personalities on the basis of which can be used to obtain legal documents and from sale of personal data to external interested organizations (a recent example of Facebook). In addition, the analysis of Internet search results for “buy EU ID-card” shows an abundance of offers to sell clones of real EU documents. Using Blockchain technology \cite{Hyperledger12} can help eliminate the first two threats and increase control over the use of such information by authorized users. Blockchain is a well known technology used in Bitcoin~\cite{Debin06,Garay02,Matsuo03} and other cryptocurrencies ~\cite{Bentov01,EOS11}. Successful attempts are also being made to introduce this technology in areas of bank transfers~\cite{Koles21}, logistics~\cite{DHL20}, energy~\cite{Kushch13}, IoT \cite{Fran14}, \cite{Kushch22} and healthcare~\cite{Kushch23}.\\
Blockchain, by definition, is a distributed database in which each subsequent block containing information is associated with the previous one. The generation of each block must be confirmed by other participants. Nowadays there are several concepts of the Blockchain consensus algorithms - Proof of Work (POW) \cite{Jakobsson04},\cite{Laurie05}, Proof of Stake (POS) \cite{Buterin09},\cite{Chohan07}, Proof of Importance (POI) \cite{Beikverdi10}, Proof of Activity (POA) \cite{Bentov08} etc. The most common algorithm is Proof of Work. However, this algorithm uses too many resources to generate blocks; it is too slow and in many cases not necessary for building a block system. For example, when all nodes of a network are well known and are state organizations. This structure of chain may be used for implementation of Blockchain in systems for checking Smart Tickets in transport, driver licenses, education degree documents etc.\\
The structure of this paper is as follows. In Section 2 we analyze the  model  of  information  system  utilizing ID-cards, ID-card structure and the structure of information in ID-card. The results obtained of the study are introduced in Section 3. Finally, we present the conclusions obtained from our research and discuss the possibilities for future work in Section 4.

\section{Background}

We further consider the methodology we propose for constructing Blockchain for the network, all nodes of which are verified servers of "Migration Police" that store personal information about citizens and existing ID-cards.
\subsection{Information system model}
\begin{figure}[h]
\centering{\includegraphics[width=100mm]{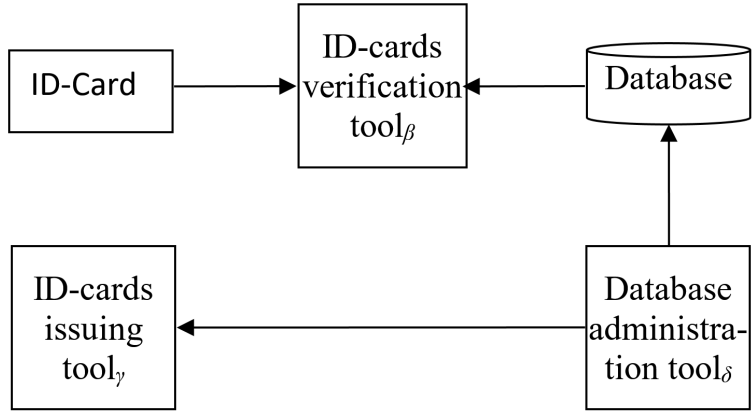}}
\caption{\textbf{The schematic representation of ID-cards verification and issuing process}\label{figure2}}
\end{figure}

The model of information system utilizing ID-cards handling consists of the following major entities:
\textit{
   \begin{itemize}
\item citizen ID-cards;
\item data of registered citizens (database in a wide sense);
\item tools for ID-cards verification;
\item tools for database administration;
\item tools for new ID-cards issuing.
 \end{itemize}
 }
The interaction between them can be presented schematically (Fig. \ref{figure2}).

It is seen from Fig. \ref{figure2} that the entities principal interaction occurs during ID-card issuing and ID-card verification. The procedure of ID-card issuing implements the following relations: 

\begin{equation*}
Issuing=
\left\{\begin{matrix}
Accounting: PD \to DB, \\
Manufacturing: PD \to ID.
\end{matrix}\right.
\end{equation*}
\\
where: $PD$ is a set of EU citizen's personal data that concerns citizenship; $DB$ is a set of database elements used by legal officers during performance of their duties; $ID$ is a set of all ID-cards issued so far.\\
It is obligatory for $ID$ and $DB$ to contain corresponding data for a given citizen. The latter is possible only in case, when these sets are isomorphic in relation to contained data $DB \Leftrightarrow ID$ . Consequently, their power should be the same $\|DB\|=\|ID\|$  and for the given $i \le \|DB\|, i \in N$  the following isomorphic equality is to take place $db_i \Leftrightarrow id_i$ , where $db_i \in DB$ and $id_i \in ID$. The latter is correct only when $DB$ and $ID$ are both particular instances of sets – lists, i.e. bounded ordered sets. 
As it was mentioned above, ID-card issuing process is an object of ID-card forgery attack, which mathematically can be described in the following form:
\begin{equation*}
\forall pd_j \in PD, Manufacturing(pd_j) \notin ID;\\
\exists db_i \in DB, \Leftrightarrow Manufacturing(pd_i).
\end{equation*}

This presentation shows that it is essential to have unattached database entry index to the ID-card. Lists could solve this issue, but their ability isn't enough, because the given $i$-th element could be replaced. That's why we suggest using data structure, which allows adding information to the DB, but forbids its deletion, replacement or alteration of Blockchain, where data from the respective ID-card would be bonded to one block.\\
The other relation shown on Fig. \ref{figure2} is ID-card verification. The process can be presented in the following form:
\begin{equation*}
Verification:ID \times DB \rightarrow \{true; false\}.
\end{equation*}

\begin{figure}[H]
\centering{\includegraphics[width=120mm]{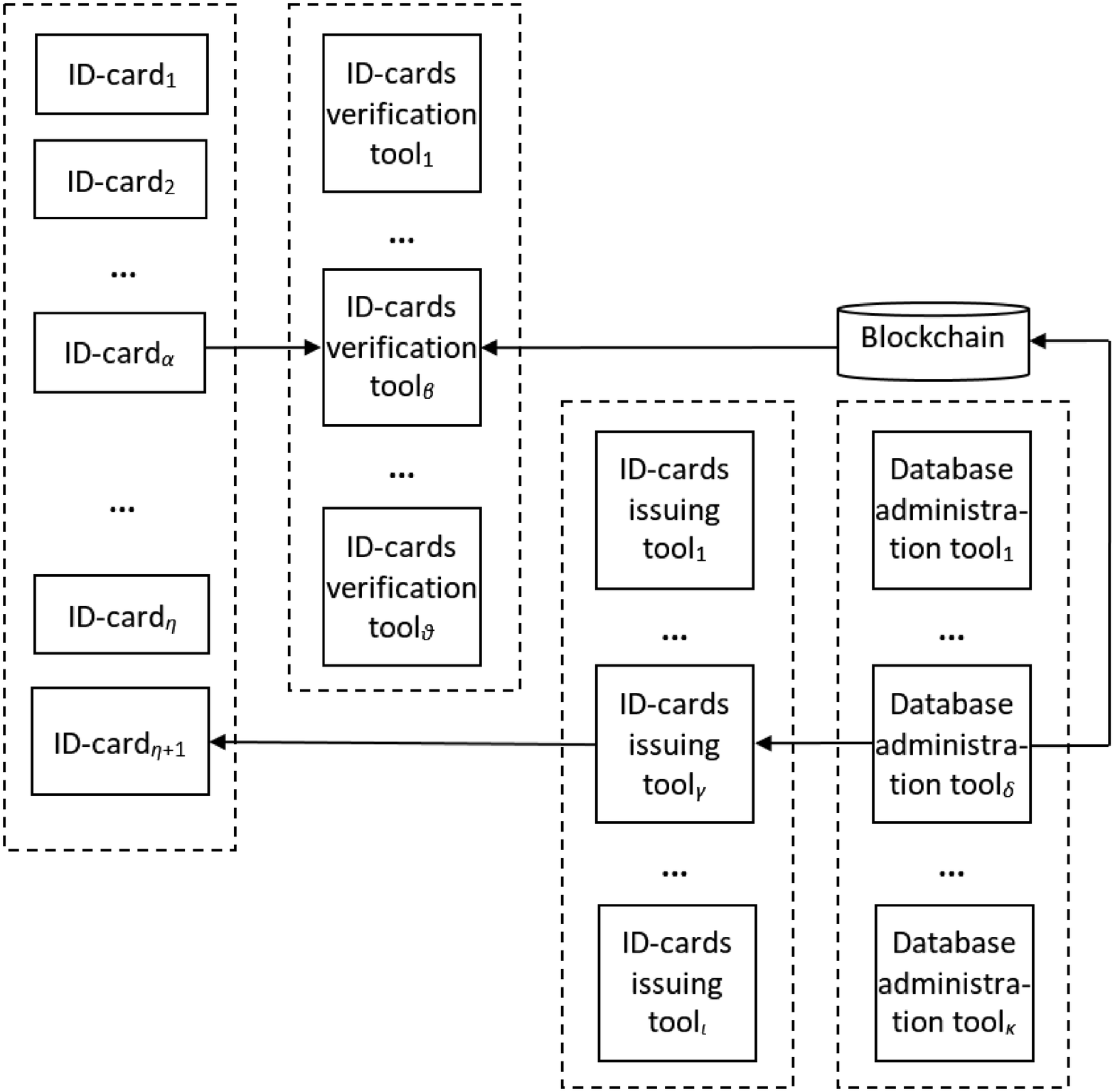}}
\caption{\textbf{Detailed schematic representation of ID-cards verification and issuing process}\label{figure3}}
\end{figure}

In the case when a Blockchain is used for data storage, the verification operator includes block number comparison, addition to the usual data content and its hash digits comparison. Moreover, fixed Blockchain structure and distribution would help to prevent attacks focused on database content.
Bearing in mind above mentioned results, the scheme presented on Fig. \ref{figure2} could be further detailed by one shown on Fig. \ref{figure3}.

\subsection{ID-card structure}
The ID-cards are used as citizen authentication factor as well as a tool for Blockchain interaction. The latter is essential in the case when Blockchain is created as shown on Fig. \ref{figure5}. In the case, ID-card user gains additional protection against intrusion into respective SubBlockchain. For instance, if intruder tries to fake his identity, he would have to break both country's security measures and signature, generated by ID-card. The signature is created by the key derived from user-specific data (e.g. fingerprints) and ID-card specific data.

To implement this feature ID-card should contain chip that provides interaction with application interface of the Blockchain. The chip needs to have the following major blocks:
\textit{
   \begin{itemize}
\item encrypted memory, which contains personal data (i.e. respective block or SubBlockchain's genesis block);
\item private key derivation block – for avoiding private key leakage;
\item signature creation block for validating changes of personal data, which could be performed when certain personal data is changed;
\item interface – for external interaction.
 \end{itemize}
 }

\begin{figure}[H]
\centering{\includegraphics[width=120mm]{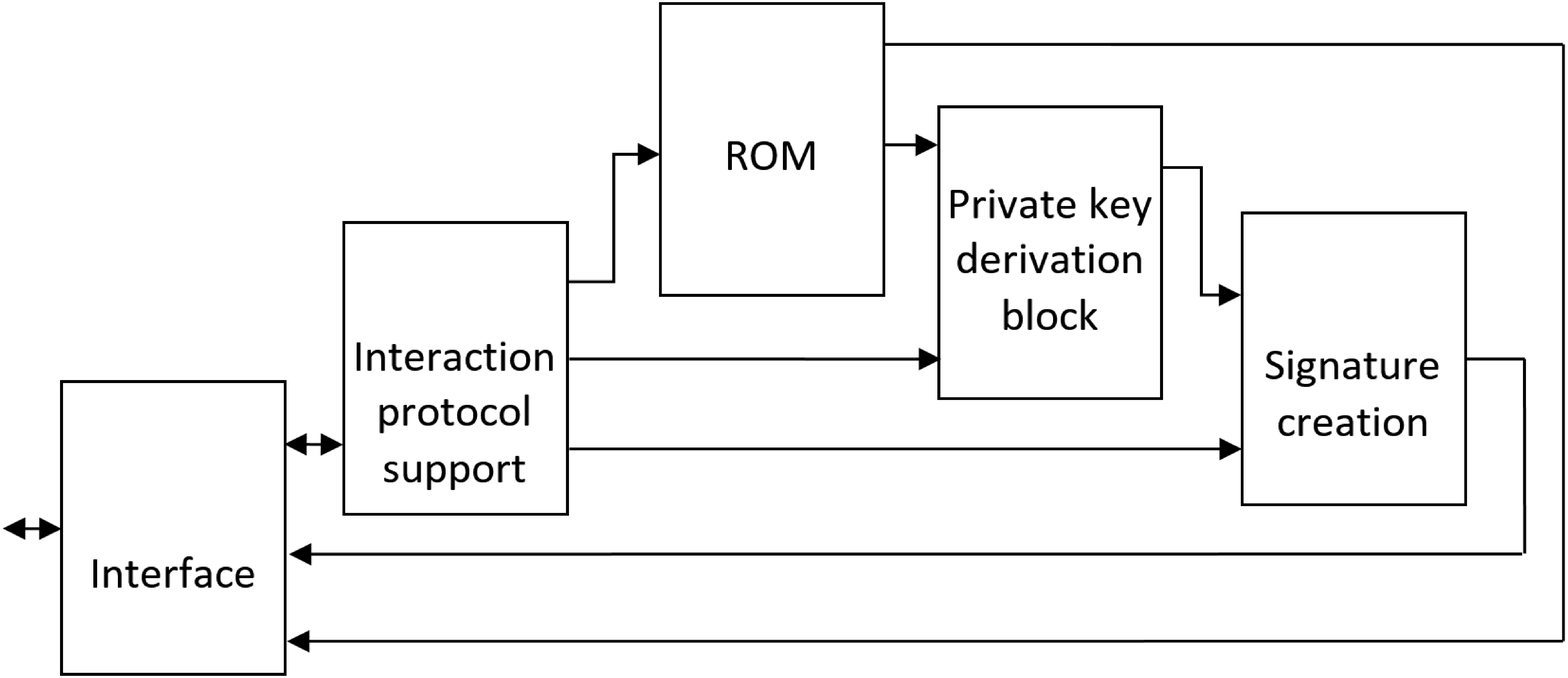}}
\caption{\textbf{Generalized structure of ID-card processor}\label{figure4}}
\end{figure}

The generalized structure of ID-card processor is shown on Fig. \ref{figure4}.

Interaction with ID-card could be performed in several ways, so certain protocol should be used. The most common task of an ID-card is providing data. Therefore, in this case ID-card would function in the following way:
\textit{
   \begin{itemize}
\item reading interface query and transmitting it to the interaction protocol support block;
\item validating the correctness of the query and determining the data fields of ROM, which are to be accessed;
\item sending respective addresses to ROM and transferring them to interface;
\item sending data according to the interface query.
 \end{itemize}
 }

Despite the comparative rarity of the cases, when data of the person is  changed,  still  in  most  cases  it shouldn't  be  performed  without  this person.  To validation of this changing the person is to possess ability to sign such transactions for instance by using the card.  For this purpose, the following steps should be performed:

\textit{
   \begin{itemize}
\item receiving respective frame-command interaction protocol support block is to wait for additional personal data input;
\item after respective data input is received interaction protocol support block transfers it to private key derivation block and sends query to respective fields of ROM for ID-card specific data;
\item using inputs private key, derivation block yields key and sends it to signature creation block;
\item meanwhile, interaction protocol support block sends data to be signed to signature creation block, where respective signature is obtained and outputted via interface.
 \end{itemize}
 }

\subsection{Block structure (Information structure on a chip)}

The block consists of the following main parts: header, data, block hash. Header structure is determined by Blockchain consensus rules and as such is heavily depend on particular protocol implementation. The most common fields of the block headers and the ones, used in particular cases are the following ones:

\textit{
   \begin{itemize}
\item version -– contains version of the block, which defines structure of other block's fields, used consensus protocol peculiarities, presumed forks markers etc;
\item hash value of previous block –- the integral part of any Blockchain, which provides cryptographic proof of its integrity;
\item hash value of block content -– generally is used for data integrity protection, but in the considered case of use it is also essential for rapid Blockchain content search;
\item creation Timestamp -– primary intent is to boost protection by limiting potential forger by this field value (its value must be in time window between the blocks preceding and following creation timestamps (Creation Timestamps)), while the field could be used to aid stored data processing;
\item creator identifier of authentication data - is supposed to determine the source of the block (node) as for legal data all sources, i.e.  nodes, are to be determined and be authorised by respective government structures.
 \end{itemize}
 }
 
In this particular case, several sources might act as creators of transaction. For instance, in case of citizenship changing, the citizen and respective governments are such source. That's why the multisig technique should be used for this type of transactions.\\
Header may include other meta-data depending on the consensus, such as nonce and difficulty for PoW consensus or number of included transactions for Blockchains where this number can vary. 
The example of EU ID-card is Estonian ID-card. The Table \ref{table1} shows the contents of a personal data  file stored on an ID-card [18].\\

\begin{table}[h!]
\caption{\textbf{Contents of a personal data  file stored on an ID-card}\label{table1}}
\begin{center}
\begin{tabular}{c c c c}
\hline
N & Content & Example & Length(bites)\\
\hline
1 & Surname & Smith & Max 28 \\
2 & First name line 1 & John & Max 15\\
3 & First name line 2 & & Max 15\\
4 & Sex	& M	& 1\\
5 & Nationality	& POL & 3\\
6 & Date of birth & 01.01.1971 & 10\\
7 & Personal ID code & 37101010021 & 11\\
8 & Document number	& X0010536  & 8 or 9\\
9 & Expiry date	& 13.08.2019 & 10\\
10 & Place of birth	& POOL/POL & Max 35\\
11 & Date of issuance & 13.08.2014 & 10\\
12 & Permit type & PERMANENT & Max 50\\
13 & Notes line 1 & EL KODANIK/EU CITIZEN & Max 50\\
14 & Notes line 2 & ALALINE ELAMISOIGUS	& Max 50\\
15 & Notes line 3 & PERMANENT RIGHT OF RESIDENCE & Max 50\\
16 & Notes line 4 & LUBATUD TOOTAD & Max 50\\
\hline
\end{tabular}
\end{center}
\end{table}

\newpage
We suggest adding extra areas in the chip memory. Thus, the data part of the block will include holder's information as it is shown in the Table \ref{table2}.
In order to provide whole block integrity we suggest using extra hashing of its content, i.e. header and data. The latter hash value is to be used to alter of the blocks.\\

\begin{table}[H]
\caption{\textbf{Contents of a personal data stored on an ID-card}\label{table2}}
\begin{center}
\begin{tabular}{c p{6cm} p{3.5cm} p{1.8cm}}
\hline
N & Content & Example & Length(bites)\\
\hline
1 & Community of issue & Milano & Max 28 \\
2 & Serial number & AA00000BB & Max 15\\
3 & First name line 1 & John	& Max 15\\
4 & First name line 2 & & Max 15\\
5 & Surname	& Smith	& Max 15\\
6 & Place of the birth & Milano & Max 35\\
7 & Sex	& M	& 1\\
8 & Nationality	& ITA & 3\\
9 & Date of birth & 01.01.1971 & 10\\
10 & Stature & 186  & 3\\
11 & Citizenship & ITA & 3\\
12 & Image of the holder's signature & "digital scan" & 50 \\
13 & Validity for expatriation & 13.08.2014 & 10\\
14 & Photography & "Digital photo"	& Max 50\\
15 & Images of 2 fingerprints (one finger of the right hand and one finger of the left hand) & "Digital photo" & Max 50\\
16 & Parents (in the case of a minor's card) & 	& Max 50\\
17 & Fiscal Code (Personal ID code) & ABCDEF00B00A111W & 16\\
18 & Address of residence & Verona str., 1, Rome & Max 50\\
19 & Tax code in the form of a bar-code & Bar-code & Max 100\\
20 & Information about the documents on the basis of which the ID-card was issued & - & Max 50\\
\hline
\end{tabular}
\end{center}
\end{table}

\begin{table*}[h!]
\caption{\textbf{Continuation of Table 2}}
\begin{center}
\begin{tabular}{c p{6cm} p{3.5cm} p{1.8cm}}
\hline
N & Content & Example & Length(bites)\\
\hline
21 & The number of the block storing information about the previous ID-card (in case of renewal or update of the card).This parameter = 0 if the card is issued for the first time & - & Max 50\\
22 & ID of migration police officer who created personal file of a citizen & - & Max 50\\
23 & Notes line 4 & Permanent residence & Max 50\\
24 & Notes line 4 & - & Max 50\\
25 & Notes line 4 & - & Max 50\\
\hline
\end{tabular}
\end{center}
\end{table*}

\section{Main results}
\subsection{Blockchain for a database}

\begin{figure}[H]
\centering{\includegraphics[width=120mm]{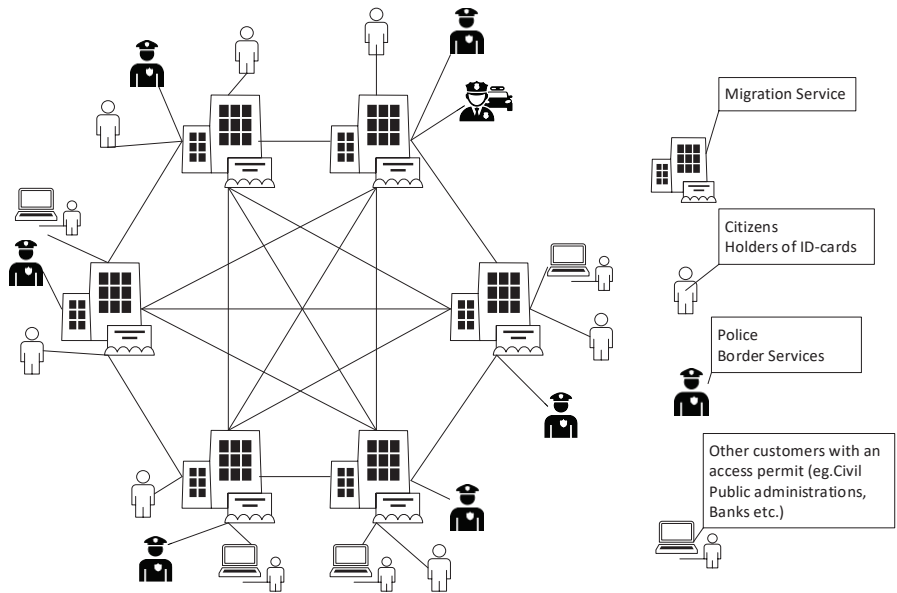}}
\caption{\textbf{Overall structure of the network. The nodes are police, special services, migration department branches, other state organizations which have access to personal data. The users of the information are police, migration officers, employees of special services  as well as third parties authorized by the state or holder of the information.}\label{figure5}}
\end{figure}

The proposed network structure is a classical peer-to-peer (P2P) topology, in which all elements are interconnected (Fig.\ref{figure5}).

\begin{figure}[H]
\centering{\includegraphics[width=100mm]{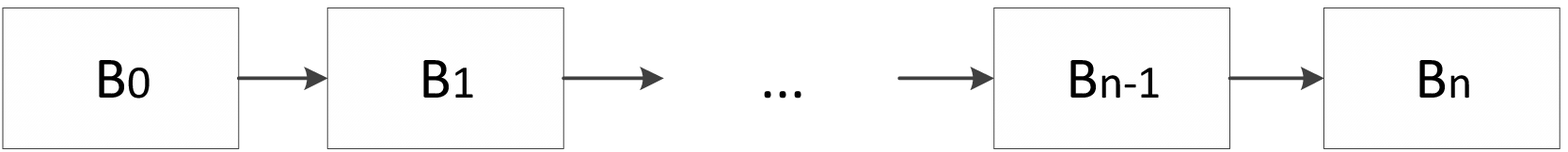}}
\caption{\textbf{General structure of Blockchain}\label{figure6}}
\end{figure}

The nodes of this network are servers of the regional branches of the migration service, which has the right to issue documents to residents (ID-cards, passports, etc.). All nodes are equal. The network has a fixed number of nodes. Each node is verified and included in the list of approved nodes. This list is stored on each node and only devices from this list can create new blocks. The structure of the chain has the form shown on Fig.\ref{figure6}.

Each block contains information about one ID-card. Transactions are personal data of ID-card holders. Created block must be confirmed by more than 50\%+1 of verified nodes. After that, information is written in the block, added to the chain and sent to other nodes. Thus, each element of the system stores a complete Blockchain. In the case of detecting a change in information in an existing block, this block will be automatically replaced with the "right" one that exists on at least 50\%+1, of other nodes of the network.
Given that the blocks in the chain are sequentially linked to each other, the alternation of one block will lead to change of the entire chain. This will be detected and corrected by the rest of the network nodes. Such an algorithm prevents the creation of fake personalities in the system with dates in the past.

\subsection{Blockchain integration}

\begin{figure}[H]
\centering{\includegraphics[width=120mm]{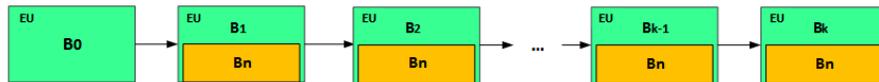}}
\caption{\textbf{Overall structure of the nested Blockchain}\label{figure7}}
\end{figure}

It should be borne in mind that the EU is building a common system for recording and controlling issued documents while building a system for protection of personal data. Therefore, it is necessary to take into account the possibility of integration of the proposed system with a system of a higher level, for example, the EU. We offer two possible options for such integration.
In the first variant, we take into account that the local Blockchain was created by each country separately and it is necessary to integrate them into a single system. To do this, it is planned to create a single EU Blockchain, which will contain blocks of local chains as transactions. In this case, the order of recording specific blocks of each country will be specified separately (Fig. \ref{figure7}).

\begin{figure}[H]
\centering{\includegraphics[width=70mm]{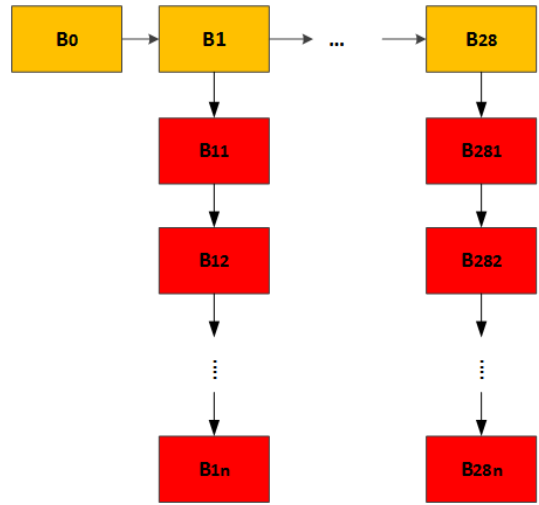}}
\caption{\textbf{Two-level Tree Blockchain for the EU countries}\label{figure8}}
\end{figure}

The second option involves creation of a more complex structure. The main chain consists of 29 blocks: B0 - genesis block and 28 (or other number) subsequent blocks (according to the number of EU member countries), each of which is a genesis block for a Subchain, which includes the Blockchain of each member country. This will allow adding new blocks to the main chain while expanding the EU in the future. Fig. \ref{figure8} shows the structure of such a Blockchain.

It should be noted, that in the case of the creation of a single system for all EU countries "from scratch", there will be no need to integrate national systems. It is enough to add a marker of the country that issued the document (the country of the creation of the block).

\subsection{SubBlockchain for access control and evaluation of ID-cards}

\begin{figure}[H]
\centering{\includegraphics[width=110mm]{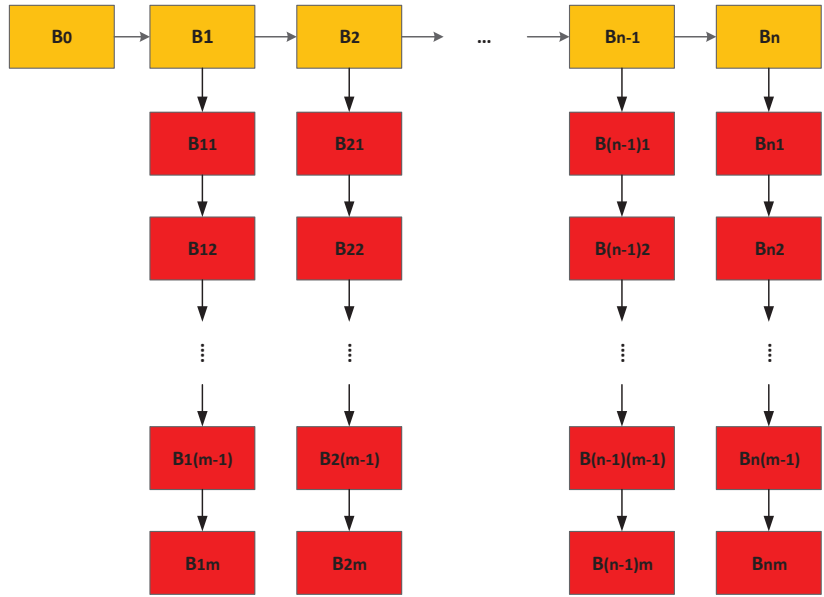}}
\caption{\textbf{Generalized structure of ID-card processor}\label{figure9}}
\end{figure}

In addition to creating fake personalities, there is also a threat of unauthorized access to personal information as well as its transfer by users to third parties. To control access, it is proposed to create a SubBlockchain, which will record information about each access attempt to a specific block and the use of information from it. The structure of such a circuit is shown on Fig. \ref{figure9}.

In this case, the block of the main chain is a Genesis Block for a Subchain. Thus, we get the number of Subchains equal to the number of blocks in the main chain. Information about access time, device from which the user entered and information about the user (as well as other necessary items for unambiguous identification of the user) will be recorded in the blocks of this chain.

\begin{figure}[H]
\centering{\includegraphics[width=115mm]{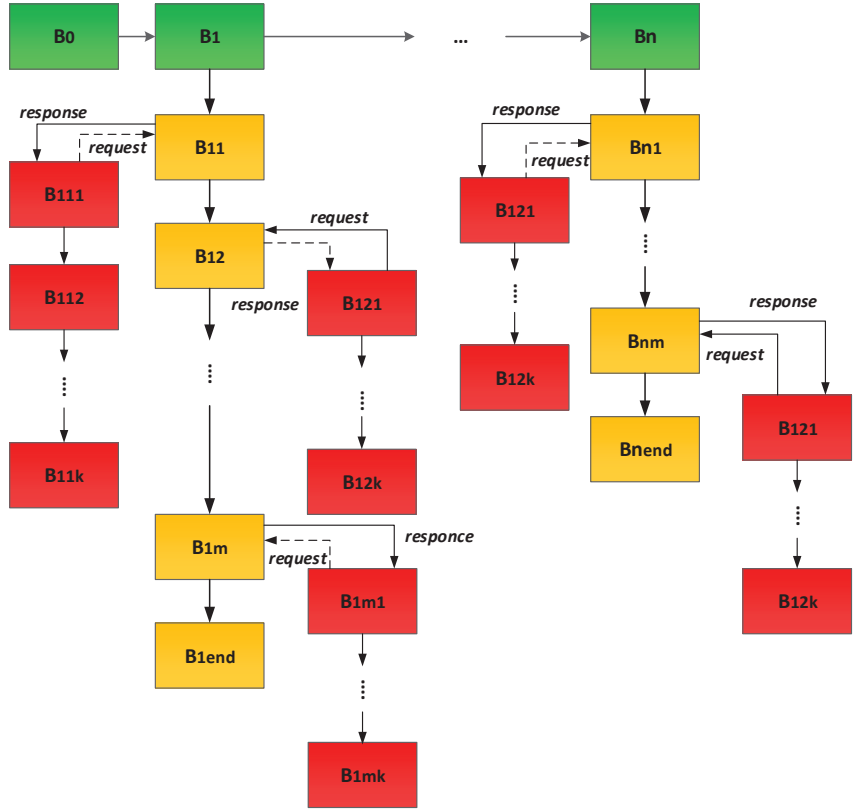}}
\caption{\textbf{Blockchain Tree. There is a system of three Blockchains which are connected which each other. The green BC is the main and contains a personal information about ID-card holder; the yellow BC is a Sub BC and contains information about reissued ID-cards; the red chain is the 2nd Sub BC and contains logs of customers, police, and other officers of special services.}\label{figure10}}
\end{figure}

If the national chain will be integrated into a higher level system as described above, the scheme for constructing the Subchain has the form shown on Fig. \ref{figure10}.
 
Specialized police scanners, smartphones with NFS module and fingerprint scanner may be used for checking the documents. In this case, the device receives HASH for the checked document from the closest node of the system and compares it with the HASH of the submitted document. Identification of the owner of the document is done by built-in fingerprint sensor. \newpage

Table 3.{\textbf{Algorithm of Block formation in a multichain}}
\begin{figure}[H]
\centering{\includegraphics[width=120mm]{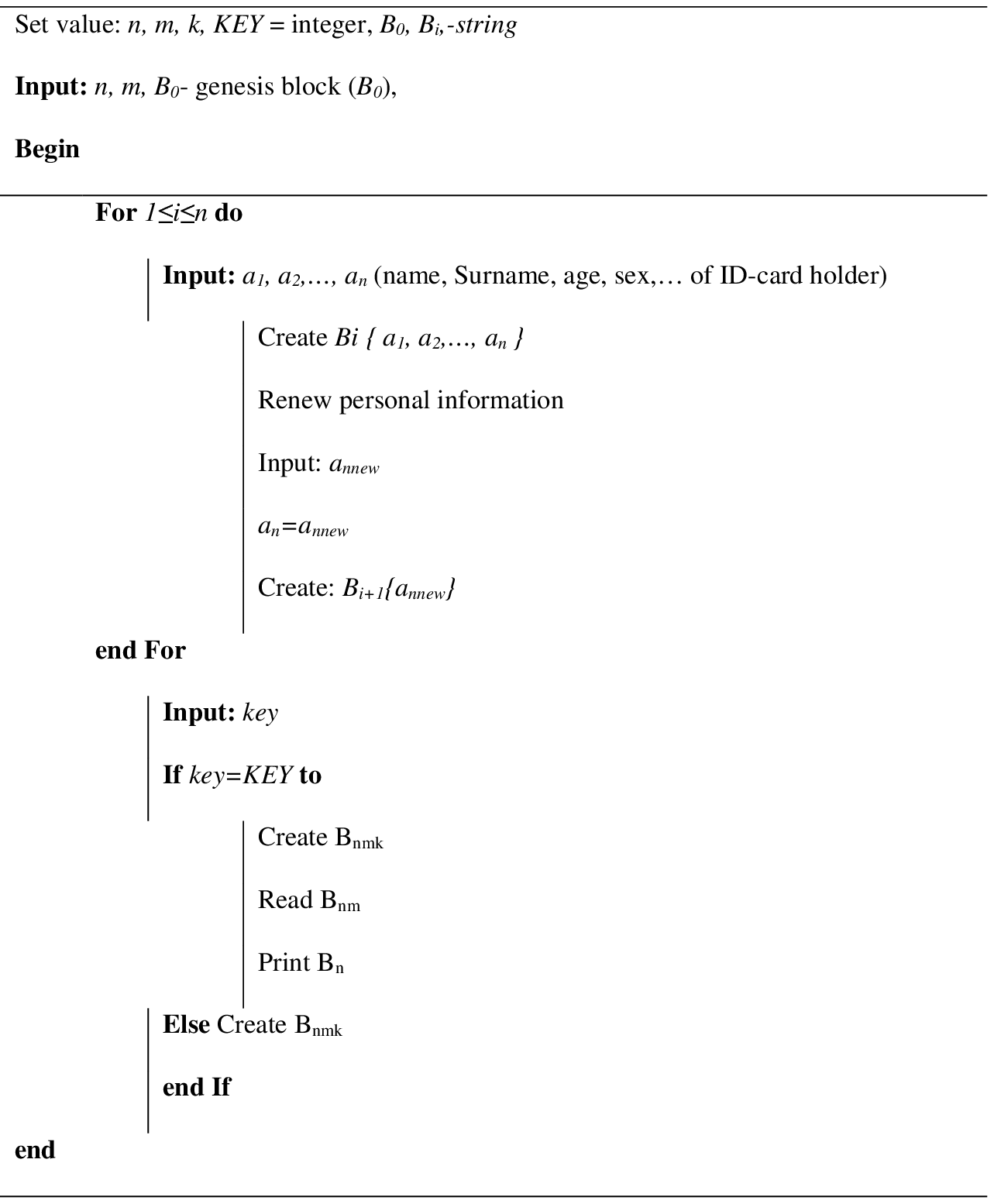}}
\end{figure}

\subsection{Mathematical model}
Mathematical structure of the main complete chain is:

\begin{equation*}
F= \bigcup_{i=0}^\infty B_i
\end{equation*}

Here $n$ - the modes number of main chain.\\
Analytic expression, describing the structure of the complete chain (Fig.\ref{figure8}), will have the form of (1).

\begin{equation}
F= B_0\bigcup_{i=1}^n B_i \bigcup_{j=1}^m B_j, [i,j \in R]
\end{equation}

where the notation is as follows: $m$ - number of elements in a Subchain; $n$ - number of elements of the main chain. \\
BC can be described by the following formula, in the case described in 3.3 of Fig.\ref{figure9}:

\begin{equation}
F= B_0\bigcup_{i=1}^n B_i \bigwedge_{j=1}^m B_j, [i,j \in R]
\end{equation}

In the case of creating an additional Subchain for the control of personal data, the formula (3) will take the following form:

\begin{equation}
F= B_0\bigcup_{i=1}^n B_i \bigwedge_{j=1}^m B_j \bigcup_{j,k}^p B_{j,k}, [i,j,k \in R]
\end{equation}

The BC structure described on Fig. 6 can be represented as follows:

\begin{equation}
F= B_0\bigcup_{i=1}^{28} B_i \bigcup_{j=1}^m B_{i,j}, [i,j \in R]
\end{equation}

here $B_0, B_i$ -- the blocks which constitute the main chain (EU chain); $B_j$ –- the blocks which are constitute the chains of each member-country. \\
Then, taking into account the second Subchain storing information about access to personal data:

\begin{equation}
F= B_0\bigcup_{i=1}^{28} B_i \bigcup_{j=1}^m B_{i,j} \bigcup_{k=1}^p B_{i,j,k}, [i,j,k \in R]
\end{equation}

\section{Summary and discussion}

In this paper we introduce a novel methodology based on Blockchain for building  storage, access control and document verification mechanisms for migration control area. The proposed work is based on Subchains, connected to a main Blockchain and each other. The solution is more security than currently existing due to use the mutual intersection of several Blockchains in the one system. This makes the process of hacking and falsification of critical information more difficult, since in the event of an attack it will be necessary to change not one but several Blockchains, which considerably increases the cost of such an attack and makes it unprofitable for the attacker.
The noted above methodology for building a storage system, access control and document verification can be used not only for ID-cards but also for other documents, such as driver's licenses, education documents, personal medical information and social security cards, etc. 
The proposed methodology still requires further improvement in order to contribute to its reliable implementation and legal compliance (in particular referring to GDPR). For example, we left out some  problems of building real networks, such as devices and communication lines delays. In our future work we will consider the problem of using various consensus algorithms with different types of Blockchains.  Also, this work did not address the use of specialized equipment for checking documents in the field with the creation of a separate closed network and using these devices as nodes of such a network. These questions will be considered and presented in the next paper.

\section{Conflicts of Interest}
The authors declare no conflicts of interest.

\section*{References}

\bibliography{mybibfile.bib}

\end{document}